\documentclass[twocolumn,superscriptaddress,showpacs,aps,amsmath,amssymb]{revtex4}
\usepackage{bm}
\usepackage{graphicx}

\newcommand{\B}{\text{\scriptsize res}}

\newcommand{\T}{{\rm total}}

\newcommand{\nl}{\nonumber \\}

\newcommand{\ep}{\epsilon}
\newcommand{\w}{\omega}

\newcommand{\be}{\begin{equation}}
\newcommand{\ee}{\end{equation}}
\newcommand{\bea}{\begin{eqnarray}}
\newcommand{\eea}{\end{eqnarray}}
\newcommand{\bsube}{\begin{subequations}}
\newcommand{\esube}{\end{subequations}}

\newcommand{\comments}[1]{}

\begin{document}

\title{Reappearance of Kondo Effect in Serially Coupled Symmetric Triple Quantum Dots}

\author{YongXi~Cheng}
\affiliation{Department of Physics, Renmin University of China, Beijing
100872, China }

\author{JianHua~Wei}\email{wjh@ruc.edu.cn}
\affiliation{Department of Physics, Renmin University of China,
Beijing 100872, China }

\author{YiJing~Yan}
\affiliation{Hefei National Laboratory for Physical Sciences at the Microscale,
University of Science and Technology of China, Hefei, Anhui 230026, China}
\affiliation{ Department of Chemistry, Hong Kong University of Science and Technology, Kowloon, Hong Kong}

\date{\today}

\begin{abstract}

We investigate the spectral properties of serially coupled triple quantum dot (TQD) system by means of the hierarchical equations of motion (HEOM) approach. We find that with the increase of the interdot coupling $t$, the first Kondo screening is followed by another Kondo effect reappearing due to the transition from the respective Kondo singlet state of individual QD to the coherence bonding state generated among the three QDs. The reappearance of Kondo effect results in the three-peak structure of the spectral functions of peripheral QD-1(3). By investigating the susceptibility $\chi$, we find that the local susceptibility of intermediate QD-2 is a positive value at weak interdot coupling, while it changes into negative value at strong interdot coupling, at which the TQD system gives rise to the reappearance of Kondo effect. We also find the slopes of $1/\chi$ will deviate from straight line behaviour at low temperature in the reappearing Kondo regime. In addition, the influence of temperature $T$ and dot-lead coupling strength $\Delta$ on the reappearing Kondo effect as well as the Kondo-correlated transport properties are afterwards exploited in detail.
\end{abstract}


\maketitle

\section{Introduction}
Triple quantum dot system as the simplest device provides an ideal platform for investigating the coded qubit, frustration and quantum teleportation \cite{2012rpp114501}. The investigation of the TQD is just the first step to study the multiple- ``impurity" configurations. More important applications of TQD are in the field of quantum computation and quantum information processing due to their extended freedom of coupling and geometric arrangement. Significantly, it leads to more interesting physics such as Fano resonances \cite{2007prb155319}, Aharonov-Bohm oscillations \cite{2008prl226810}, Quantum phase transitions \cite{2006prb045312} and Kondo effect \cite{2006prb153307,2013prl047201}.

Recently, TQD has been paid much attention from both experimental and theoretical aspects. Gaudreau \emph{ et al.} \cite{2006prl036807} create the TQDs device for the first time from a two-dimensional electron gas by applying suitable gate voltages. The stability diagram of the few-electron triple dots system is mapped out experimentally with charge detection technology. Then, the serial triple QDs \cite{2007prb075306} electrostatically defined in a GaAs/AlGaAs heterostructure and the three-dimensional nature of the stability diagram\cite{2010prb075304} is presented. A tunable few-electron lateral TQD arranged in series is designed in 2009 \cite{2009apl193101} and the charge stability diagrams of this device for different electrons and two configurations are showed by using one of the quantum point contacts as a charge detector. Moreover, the QDs are laterally coupled in a linear or triangular geometric arrangements \cite{2008prb193306,2008apl013126,2008pe1656} and a nearby quantum point contact is employed as an electrometer to probe the stability diagram. An equilateral triangular arrangement \cite{2008pe1322,2009apl092103} and collinear arrangement of TQD  \cite{2010pe899} not using quantum point contact electrometers are also designed on $\mathrm{GaAs/Al_{0.3}Ga_{0.7}As/GaAs/Al_{0.3}Ga_{0.7}As/GaAs}$ double barrier resonant tunneling structure.

At low temperature, the Kondo effect exhibits in the nanoscale Coulomb blockade systems with degenerate ground states. As a many-body phenomenon, the Kondo effect in the QDs system acquires great interest as a significant role in contributing to the quantum transport properties of multiple - ``impurity" configurations. It has been widely studied theoretically and experimentally in the single and double QDs systems. But the study of Kondo physics for the TQD is a great challenge for theorists due to the difficulties and accuracies of the present methods, especially, when the fourth-operator term of Hamiltonian is taken into account. Despite the difficulties of the theoretical study of TQD, there is still some work in literature. Fermi-Liquid versus non-Fermi-Liquid behavior in TQDs coupled in series are discussed using the approach of numerical renormalization group(NRG) \cite{2007prl047203}. Two-channel Kondo physics appears in serially coupled TQD with odd electron occupation \cite{2005el218}. In a wide temperature interval, the system exhibits the two-channel Kondo effect and has non-Fermi-Liquid property for the appropriate hopping parameters. Kondo phases resulted from the transition from local-moment to molecular-orbital behavior are presented in a zero-temperature phase diagram with different hopping matrix elements \cite{2006prb153307}. Zhan-tan Jiang \emph{et al.} \cite{2005prb045332} theoretically studied the equilibrium and nonequilibrium Kondo transport properties of the serially coupled TQD by means of slave-boson mean-field approximation approach. But the structure is only on the symmetrical TQD system with no Coulomb interaction in the intermediate QD, as slave-boson mean-field method might not be available for the appearance of the finite Coulomb interaction in intermediate QD. Other literature focus on the Kondo effect and spectral properties of the different structures, such as mirror symmetry TQD \cite{2013prl047201,2006prb235310,2010prb165304}, triangular TQD \cite{2009prb155330,2010prb115330,2011prb205304,2013prb035135} and parallel TQD \cite{2007prb115114}. In general, the Kondo effect of the TQD system has not been systematically studied and a comprehensive picture is missing.

Here, we characterize the spectral properties and Kondo transport in serially coupled symmetric TQD systems by means of hierarchical equations of motion (HEOM) approach\cite{Jin08234703,2008jcp184112,2009jcp124508,2008njp093016,2009jcp164708} established based on the Feynman-Vernon path-integral formalism, in which all the system-bath correlations are taken into consideration. In this paper, the Kondo effect of the TQD with some intriguing properties which are not presented in single or double QDs systems is explored. We find that Kondo effect reappears in the TQD when appropriate requirement is satisfied, and thus investigate the thermodynamic properties and dynamical properties in the Kondo regime.

The paper is organized as follows. In Sec.II the TQD model adopted in this work is introduced and the principal features of HEOM approach are briefly reviewed. In Sec.III the spectral properties and susceptibility of TQD are investigated and some intriguing performances of the reappearing Kondo effect are revealed. Then, the current through the serially coupled TQD corresponding to the reappearing Kondo effect are explored. Finally, conclusions is given in Sec.IV.

\section{MODEL AND HEOM APPROACH}
A serially coupled TQD system is the model we study (see figure 1(a)). The two peripheral quantum dots (QD-1 and QD-3) are directly coupled to the leads, while the intermediate QD (QD-2) is not directly coupled with the leads. The localized QDs constitute the open system of primary interest, and the surrounding reservoirs of itinerant electrons are treated as environment. The total Hamiltonian for the system is $H=H_{dots}+H_{leads}+H_{coupling}$, where the interacting TQD
\begin{align}\label{hs2}
   H_{dots}=\sum_{\sigma i=1,2,3}[\epsilon_{i\sigma}\hat{a}^\dag_{i\sigma}\hat{a}_{i\sigma} + U_{i} n_{i\sigma}n_{i\bar{\sigma}}]
 \nl    + \sum_{\sigma}(t_{12}\hat{a}^\dag_{1\sigma}\hat{a}_{2\sigma}+t_{23}\hat{a}^\dag_{2\sigma}\hat{a}_{3\sigma}+\text{H.c.})
  \end{align}

here $\hat{a}_{i\sigma}^\dag$ ($\hat{a}_{i\sigma}$) is the operator that creates (annihilates) a spin-$\sigma$ electron with energy $\epsilon_{i\sigma}$ ($i=1,2,3$) in the dot $i$. $n_{i\sigma}=\hat{a}^\dag_{i\sigma}\hat{a}_{i\sigma}$ corresponds to the operator for the electron number of dot $i$. $U_{i}$ is the on-dot Coulomb interaction between electrons with spin $\sigma$ and $\bar{\sigma}$ (opposite spin of $\sigma$), while $t_{12}$ ($t_{23}$) is the interdot coupling strengths between the QD-1(3) and QD-2, determined by overlapping integral of them. For simplicity, we will take $t_{12}=t_{23}=t$ in our model.

In what follows, the symbol $\mu$ is adopted to denote the electron orbital (including spin, space, \emph{etc.}) in the system for brevity, i.e.,  $\mu=\{{\sigma},i...\}$. The device leads are treated as noninteracting electron reservoirs and the Hamiltonian can be written as
$H_{leads}=\sum_{k\mu\alpha=L,R}\epsilon_{k\alpha}\hat{d}^\dag_{k\mu\alpha}\hat{d}_{k\mu\alpha}$, the term of dot-lead coupling is $H_{coupling}=\sum_{k\mu\alpha}t_{k\mu\alpha}\hat{a}^\dag_{\mu}\hat{d}_{k\mu\alpha}+\text{H.c.}$, with $\epsilon_{k\alpha}$ being the energy of an electron with wave vector $k$ in the $\alpha$ lead, and $\hat{d}^\dag_{k\mu\alpha}$($\hat{d}_{k\mu\alpha}$) corresponding creation (annihilation) operator for an electron with the $\alpha$-reservoir state $|k\rangle$ of energy $\epsilon_{k\alpha}$. To describe the stochastic nature of the transfer coupling, it can be written in the reservoir $H_{leads}$-interaction picture as $H_{coupling}=\sum_{\mu}[f^\dag_{\mu}(t)\hat{a}_{\mu} +\hat{a}^\dag_{\mu}f_{\mu}(t)]$, with  $f^\dag_{\mu}=e^{iH_{leads}t}[\sum_{k\alpha}t^{*}_{k\mu\alpha}\hat{d}^\dag_{k\mu\alpha}]e^{-iH_{leads}t}$ being the stochastic interactional operator and satisfying the Gauss statistics. Here, $t_{k\mu\alpha}$ denotes the transfer coupling matrix element. The influence of electron reservoirs on the dots is taken into
account through the hybridization functions, which is assumed Lorentzian form, $\Delta_{\alpha}(\w)\equiv\pi\sum_{k} t_{\alpha k\mu}t^\ast_{\alpha
k\mu} \delta(\w-\ep_{k\alpha})=\Delta W^{2}/[2(\w-\mu_{\alpha})^{2}+W^{2}]$, where $\Delta$ is the effective quantum dot-lead coupling strength, $W$ is the band width, and $\mu_{\alpha}$ is the chemical potentials of the $\alpha$ ($\alpha=L,R$) lead \cite{2012prl266403,2015njp033009,2013zheng086601}.

In this paper, a hierarchical equations of motion approach (HEOM) developed in recent years is employed to study the TQD system. The HEOM based numerical approach is potentially useful for addressing the interacting strong correlation systems and has been employed to study dynamic properties, such as the dynamic Coulomb blockade Kondo, dynamic Kondo memory phenomena and time-dependent transport with Kondo resonance in QDs systems \cite{2008njp093016,2009jcp164708,2015njp033009}. The resulting hierarchical equations of motion formalism are in principle exact and applicable to arbitrary electronic systems, including Coulomb interactions, under the influence of arbitrary time-dependent applied bias voltage and external fields. The outstanding issue of characterizing both equilibrium and nonequilibrium properties of a general open quantum system are referred to in Refs. \cite{Jin08234703,2012prl266403,Zhe121129,2015njp033009}. It is essential to adopt appropriate truncated level to close the coupled equations. The numerical results are considered to be quantitatively accurate with increasing truncated level and converge. It has been demonstrated that the HEOM approach achieves the same level of accuracy as the latest high-level numerical renormalization group and time-dependent density-matrix renormalization group methods for the prediction of various dynamical transport properties at equilibrium and nonequilibrium \cite{2012prl266403,2015njp033009}.

The HEOM theory established based on the Feynman-Vernon path-integral formalism adopts a general form of the system Hamiltonian, in which all the system-bath correlations are taken into consideration. The different transport processes can be handled in a unified manner. It is applicable to a wide range of system parameters without additional derivation and programming efforts and can characterize both static and transient electronic properties of strongly correlated system \cite{2015njp033009}. The reduced density matrix of the quantum dots system $\rho^{(0)}(t) \equiv {\rm tr}_{\B}\,\rho_{\T}(t)$ and a set of auxiliary density matrices  $\{\rho^{(n)}_{j_1\cdots j_n}(t); n=1,\cdots,L\}$ are the basic variables in HEOM. Here $L$ denotes the terminal or truncated tier level. The HEOM that governs the dynamics of open system assumes the form of \cite{Jin08234703,2012prl266403}:
\begin{align}\label{HEOM}
   \dot\rho^{(n)}_{j_1\cdots j_n} =& -\Big(i{\cal L} + \sum_{r=1}^n \gamma_{j_r}\Big)\rho^{(n)}_{j_1\cdots j_n}
     -i \sum_{j}\!     
     {\cal A}_{\bar j}\, \rho^{(n+1)}_{j_1\cdots j_nj}
\nl &
    -i \sum_{r=1}^{n}(-)^{n-r}\, {\cal C}_{j_r}\,
     \rho^{(n-1)}_{j_1\cdots j_{r-1}j_{r+1}\cdots j_n}
\end{align}
the $n$th-order auxiliary density operator $\rho^{(n)}$ can be defined via auxiliary influence functional as  $\rho^{(n)}_{\textbf{j}}(t)\equiv \mathcal{U}^{(n)}_{\textbf{j}}(t,t_{0})\rho(t_{0})$, with the reduced Liouville-space propagator $ \mathcal{U}^{(n)}_{\textbf{j}}(\psi,t;\psi_{0},t_{0})$ referred to in \cite{Jin08234703}.

We adopt the index $\textbf{j}=\{j_1\cdots j_n\}$ and $\textbf{j}_{r}=\{j_1\cdots j_{r-1}j_{r+1}\cdots j_n\}$. The action of superoperators respectively is
\begin{align}\label{HEOM}
   {\cal A}_{\bar j}\, \rho^{(n+1)}_{\textbf{j}j} = a^{\bar{o}}_{\mu} \rho^{(n+1)}_{\textbf{j}j}
+(-)^{n+1} \rho^{(n+1)}_{\textbf{j}j}a^{\bar{o}}_{\mu}
\end{align}
\begin{align}\label{HEOM}
  {\cal C}_{j_r}\,\rho^{(n-1)}_{\textbf{j}_{r}} = \sum_{\nu} \{{\cal C}^{o}_{\alpha\mu\nu} a^{o}_{\nu}\rho^{(n-1)}_{\textbf{j}_{r}}
 &
    -(-)^{n-1}\, {\cal C}^{\bar{o}}_{\alpha\nu\mu}\,\rho^{(n-1)}_{\textbf{j}_{r}}a^{o}_{\nu}\}
\end{align}
where, the index $j \equiv (o\mu m)$ denotes the transfer of an electron to/from ($o=+/-$) the impurity state $|\sigma\rangle$, associated with the characteristic memory time $\gamma_m^{-1}$. The total number of distinct $j$ indexes involved is determined by the preset level of accuracy for decomposing reservoir correlation functions by exponential functions. $a_{\mu}^{o}$ ($a_{\mu}^{\bar{o}}$) corresponds the creation (annihilation) operator for an electron with the $\mu$ electron orbital (including spin, space, \emph{etc.}). The correlation function ${\cal C}^{o}_{\alpha\mu\nu}(t-\tau)=\langle
f^{o}_{\alpha\mu}(t)f^{\bar{o}}_{\alpha\nu}(\tau)\rangle_{B}$ follows immediately the time-reversal symmetry and detailed-balance relations.

The spectral function $A(\omega)$ which exhibits prominent Kondo signatures at low temperatures can be evaluated in two ways: either with a time-domain scheme or by calculations in the frequency domain. The time-domain scheme starts with the evaluation of system correlation functions $\tilde{\cal C}_{\hat{a}^\dag_{\mu}\hat{a}_{\mu}}(t)$ and $\tilde{\cal C}_{\hat{a}_{\mu}\hat{a}^\dag_{\mu}}(t)$ via the time evolution of the HEOM propagator $\hat{\mathcal{G}}_{eq}(t)$ at $t > 0$. The spectral function obtained straightforwardly by a half Fourier transform is
\begin{align}\label{hd}
    A_{\mu}(\omega)=\frac{1}{\pi}Re\Big\{\int^{\infty}_{0}dt \{\tilde{\cal C}_{\hat{a}^\dag_{\mu}\hat{a}_{\mu}}(t)+
    [\tilde{\cal C}_{\hat{a}_{\mu}\hat{a}^\dag_{\mu}}(t)]^{\ast}\}e^{i\omega t}\Big\}
  \end{align}%
The frequency domain scheme involves the half Fourier transform of HEOM to evaluate $\bar{\cal C}_{AB}(\omega)=\int^{\infty}_{0}dt{\cal C}_{AB}(t)e^{i\omega t}$. The spectral function is then
\begin{align}\label{hd}
    A_{\mu}(\omega)=\frac{1}{\pi}Re[\bar{\cal C}_{\hat{a}_{\mu}\hat{a}^\dag_{\mu}}(\omega)+\bar{\cal C}_{\hat{a}^\dag_{\mu}\hat{a}_{\mu}}(-\omega)]
  \end{align}%
The total system spectral function is $A(\omega)=\Sigma^{N}_{\mu=1}A_{\mu}(\omega)$. The details of the HEOM formalism and the derivation of the spectral function of system are supplied in the Refs.\cite{Jin08234703,2008njp093016,2009jcp164708,2012prl266403}. Consequently, the varied physical quantities such as the local magnetic susceptibility and current can be acquired via the HEOM-space linear response theory.

\section{RESULTS AND DISCUSSION}

We present the numerical solution of the TQD model in figure 1(a) using the HEOM method. For simplicity, the three QDs in this model are assumed equivalent and possesses the electron-hole symmetry. So we will take the same parameters for the three QDs in calculations as follow : $\epsilon_{i}(i=1,2,3) = -1.0\mathrm{meV}$ and $U_{i} (i=1,2,3) = 2.0\mathrm{meV}$. In figure 1(b) we show the single-dot spectral function of the TQD system $A(\omega)$, as a function of frequency $\omega$ for different values of interdot coupling strengths $t$. The quantum dot-lead coupling strength is $\Delta = 0.2\mathrm{meV}$, the band width is $W = 2.0\mathrm{meV}$ and temperature is $K_{B}T = 0.03\mathrm{meV}$. We note that for weak interdot coupling strength ($t = 0.05\mathrm{meV}$), the spectral function both for QD-1(3) and QD-2 exhibit a single Kondo peak centered at $\omega = 0$. A continuous transition is observed from the Kondo state exhibiting a single-peak Kondo resonance to another exhibiting a double peak by increasing the interdot coupling strength $t$. The most interesting issue in TQD system is that the Kondo effect reappears at the strong interdot coupling strength for the QD-1(3) (figure 1(b)), accompanied by a three-peak structure of the spectral function. The width of the central Kondo peak ($\omega = 0$) broadens and the height increasing with $t$. On the contrary, the Kondo peak of the spectral function for QD-2 disappears(inset of figure 1(b)) with the increase of the interdot coupling strength.
\begin{figure}
\begin{minipage}[b]{0.5\textwidth}
\includegraphics[width=0.8\textwidth]{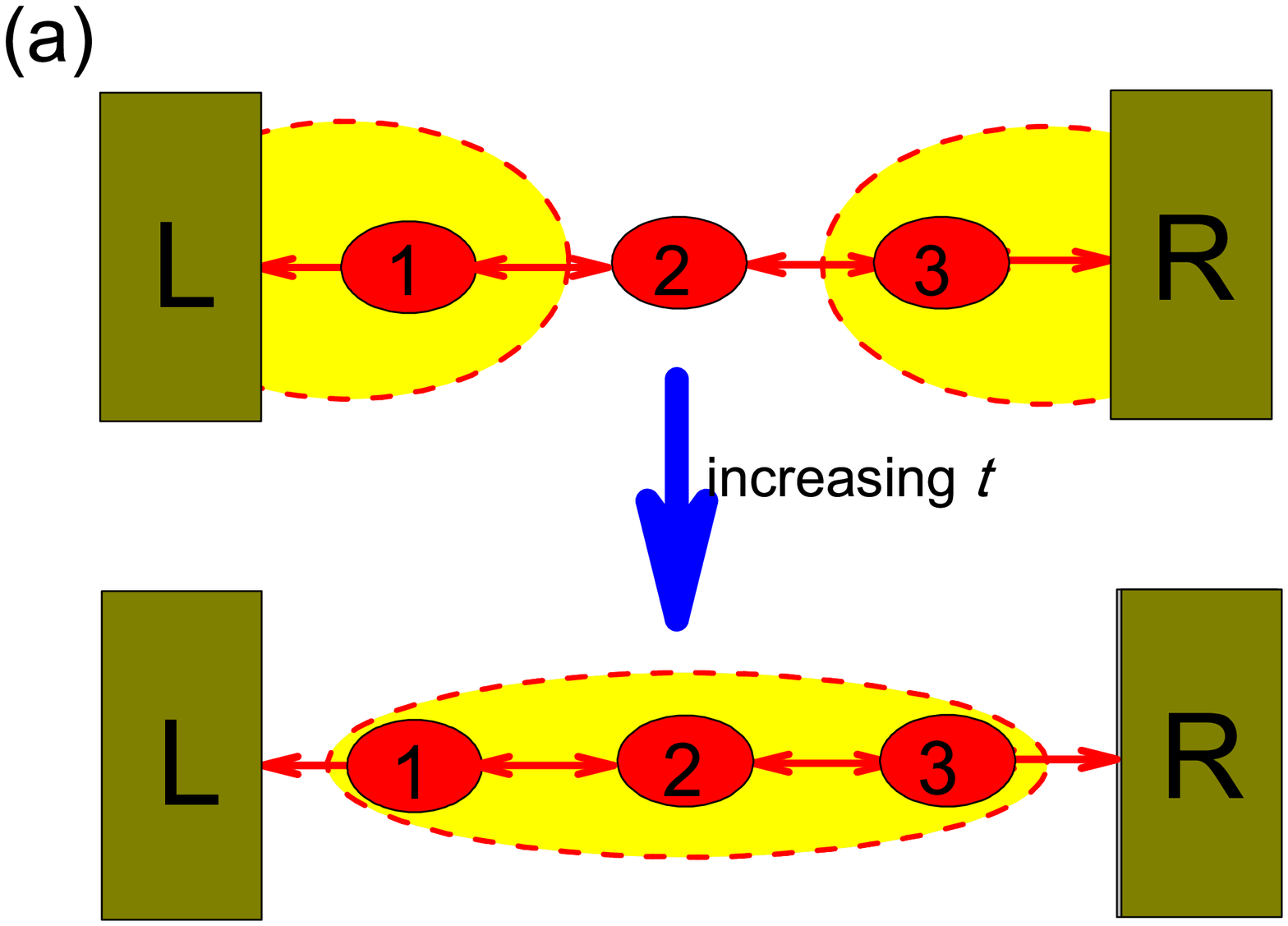}
\label{fig1(a)}
\end{minipage}%

\begin{minipage}[b]{0.5\textwidth}
\includegraphics[width=1\textwidth]{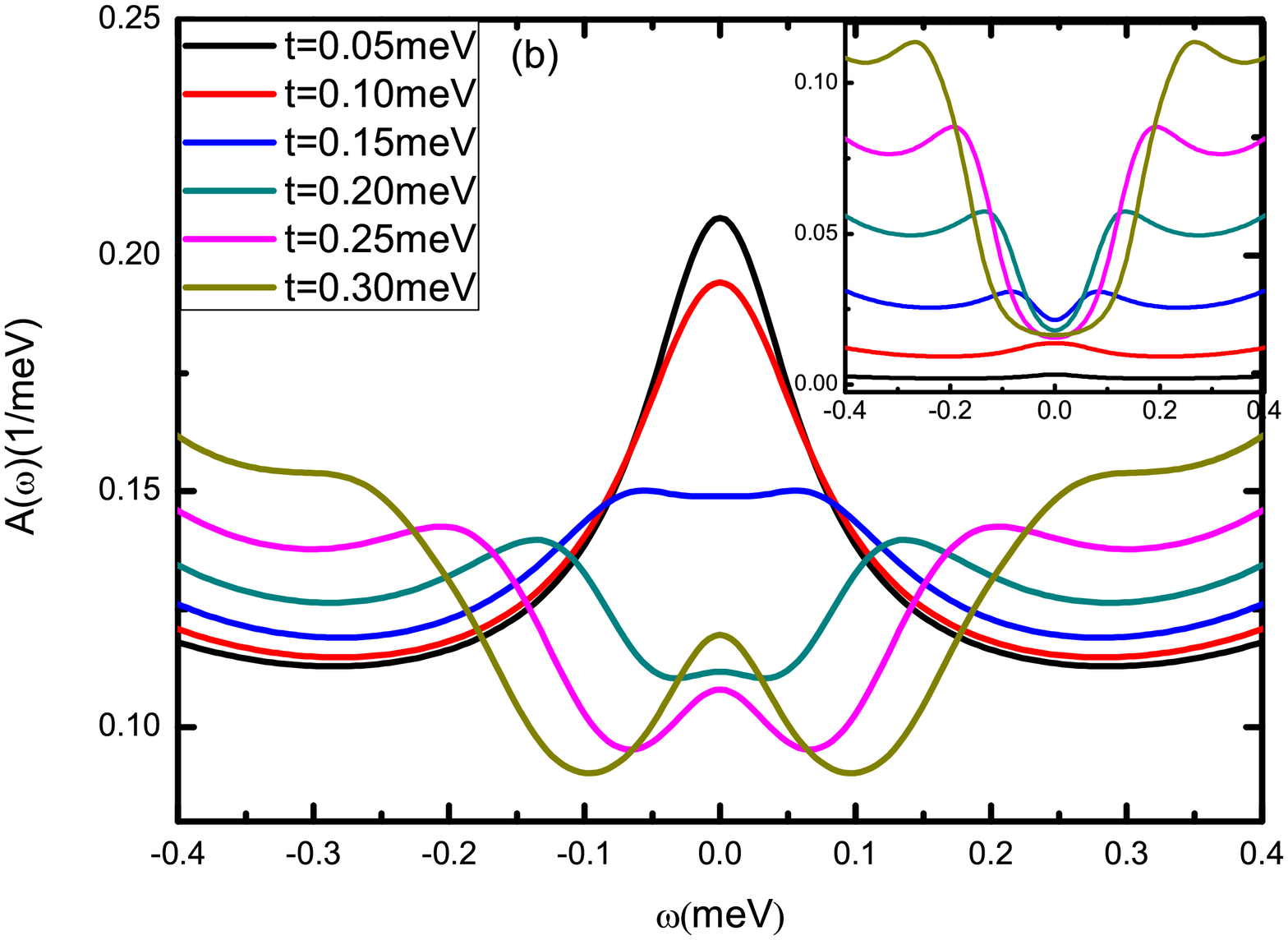}

\label{fig1(b)}
\end{minipage}

\caption{(Color online) (a) Schematic description of serially coupled TQD device. With increasing interdot coupling strength $t$, a transition from the respective Kondo singlet state of individual QD to the coherence bonding state generated between the three QDs. (b) The spectral function $A(\omega)$ of the QD-1(3) in TQD system with different interdot coupling strengths $t$ calculated by the HEOM approach. The inset of (b) shows the spectral function $A(\omega)$ of QD-2. The parameters adopted are $\epsilon_{i}(i=1,2,3) = -1.0\mathrm{meV}$ and $U_{i}(i=1,2,3) = 2.0\mathrm{meV}$, $W = 2.0\mathrm{meV}$, $\Delta = 0.2\mathrm{meV}$, $K_{B}T = 0.03\mathrm{meV}$.}
\label{fig1}
\end{figure}

At small interdot coupling strength, such as $t = 0.05\mathrm{meV}$, the quantum dot-lead coupling strength is much large ($\Delta = 0.2\mathrm{meV}$), QD-1(3) prefers to form its respective Kondo singlet with the delocalized electrons of leads (upper part of figure 1(a)), with the result that the QD-1(3) moment screened by the left (right) lead is dominant and the spectral function shows similar behavior to the single QD system. Just as the Kondo state of single QD, a single peak Kondo resonance emerges on the QD-1(3). Meanwhile, the left (right) lead and QD-1(3) constitute an effective surrounding reservoir to QD-2. A lower single Kondo peak is observed on QD-2 due to the weak interdot coupling strength between QD-2 and QD-1(3). For example, the height of the Kondo peak for QD-1(3) is $0.19\mathrm{meV}^{-1}$ at $t = 0.10\mathrm{meV}$, while it possesses only $0.02\mathrm{meV}^{-1}$ for QD-2. With the increase of the interdot coupling strength, the effective antiferromagnetic exchange interaction $J_{eff} = 4t^{2}/U$ causes the single Kondo peak both for the three QDs splitting to two peaks with the position at $\omega \sim \pm J_{eff}$. The splitting of the Kondo peak originating from the interdot coupling strength indicates the quantum coherence between the many-body Kondo states on each dot. If the interdot coupling strength being further increased, the spin of the adjacent QDs are beginning to form an antiparallel arrangement between each other due to the antiferromagnetic exchange interaction $J_{eff}$. The coherence bonding state ($|\uparrow,\downarrow,\uparrow \rangle$ or $|\downarrow,\uparrow,\downarrow\rangle$) of the TQD system persists in a local moment phase (lower part of figure 1(a)). This state can also be screened by the conduction electrons of leads. For this reason, first appearance Kondo screening followed by a reappearing Kondo effect as a three-peak structure of spectral function is observed in QD-1(3). As for QD-2, the screened QD-1 and QD-3 can be considered a equivalent QD, with the result that the QD-2 exhibits a double peaked behaviour closely analogous to that of the serially coupled double QDs (inset of figure 1(b)). Significantly, we note that the transition of spectral function for QD-1(3) is smooth (crossover) and there is no abrupt phase transition. The behavior of this reappearing Kondo effect of TQD system possesses some intriguing properties which are not presented in single or double QDs systems.

\begin{figure}
\begin{minipage}[b]{0.5\textwidth}
\includegraphics[width=1\textwidth]{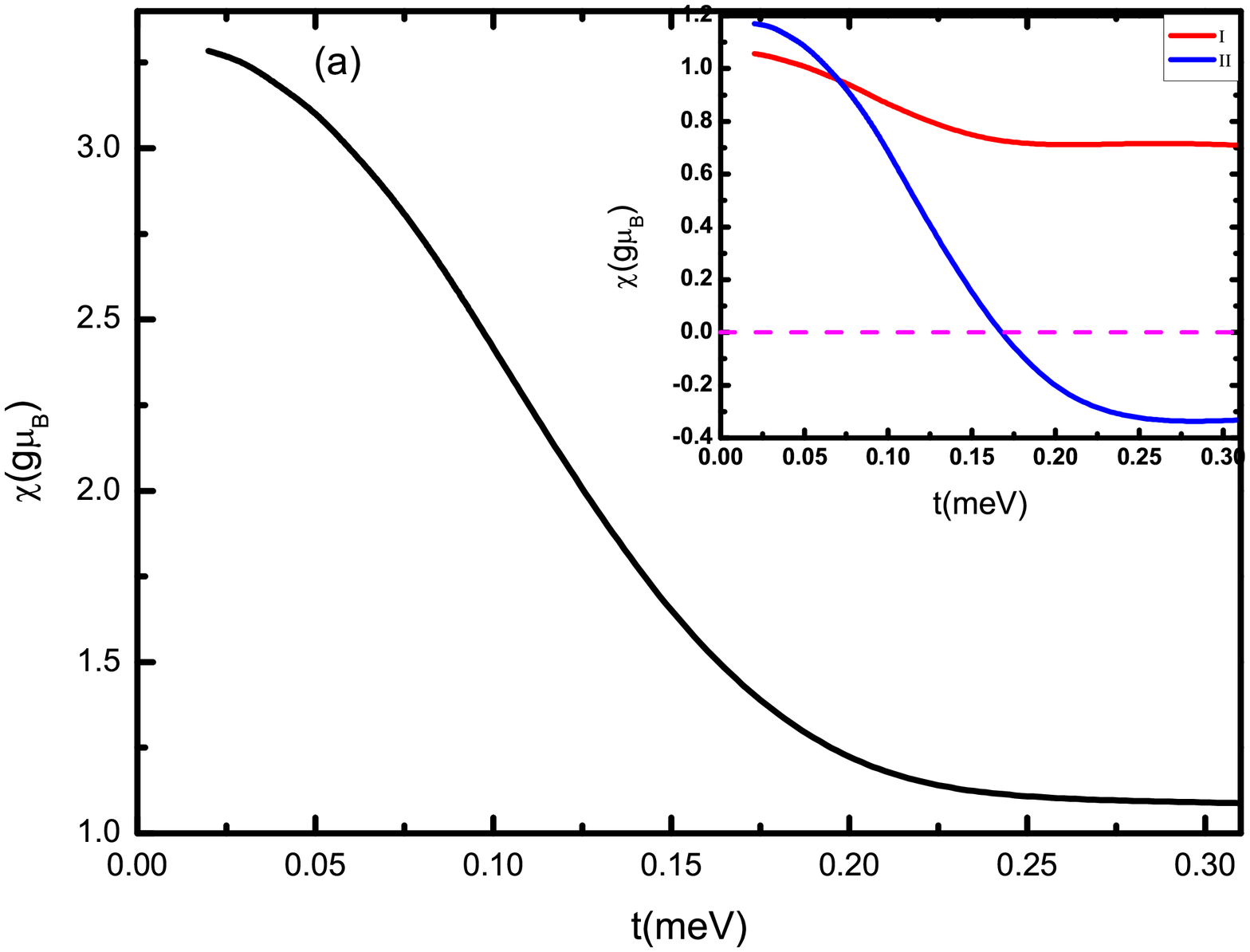}
\label{fig2(a)}
\end{minipage}%

\begin{minipage}[b]{0.5\textwidth}
\includegraphics[width=1\textwidth]{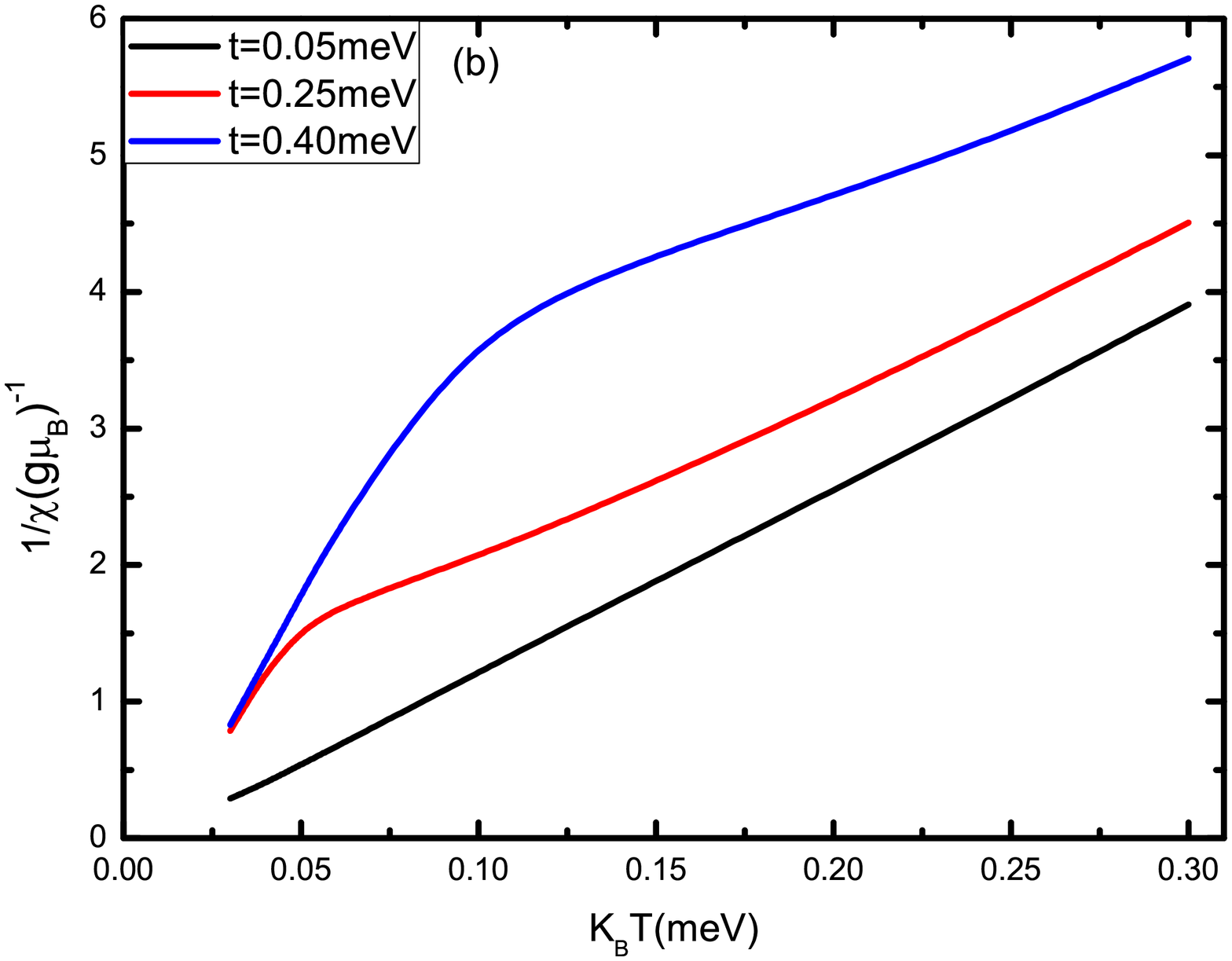}

\label{fig2(b)}
\end{minipage}

\caption{(Color online) (a) The susceptibility of the TQD system versus interdot coupling strength $t$. The inset shows the local susceptibility of the QD-1(3) (I) and of the QD-2 (II). (b) The susceptibility of the TQD system $1/\chi$ versus temperatures $K_{B}T$ with different interdot coupling strengths $t$. The parameters adopted are $\epsilon_{i}(i=1,2,3) = -1.0\mathrm{meV}$ and $U_{i}(i=1,2,3) = 2.0\mathrm{meV}$, $W = 2.0\mathrm{meV}$, $K_{B}T = 0.03\mathrm{meV}$.} \label{fig2}
\end{figure}

To further elucidate the physical mechanism of the reappearing Kondo effect, we then study the characters of the susceptibility. The susceptibility of the TQD system versus interdot coupling strength $t$ is studied in figure 2(a). By comparison, the parameters adopted are the same as figure 1(b). Firstly, the susceptibility of TQD system decreases monotonically with interdot coupling strength $t$. At very strong interdot coupling, there is no local moment of TQD system. The QDs spins are completely screened by conduction electrons with the result of a finite value of susceptibility. Similarly, the local susceptibilities both for QD-1(3) (curve (I)) and QD-2 (curve (II)) showed in the inset of figure 2(a) share the same transition behaviour to the total susceptibility of the TQD system. Importantly, the local susceptibility of QD-2 is extremely sensitive to $t$ than QD-1(3). One interesting issue is the transition of local susceptibility of QD-2 induced by $t$, from the positive value at weak interdot coupling to the negative value at strong interdot coupling. This behavior also demonstrates the Kondo physics of TQD depicted in figure 1(a): at weak interdot coupling strength, it is predominant that the Kondo singlet state of the QD-1(3) forming with the itinerant electron of the left(right) lead, and the QD-2 approaches to an isolated particle corresponding to a free spin, as sketched in the upper part of figure 1(a). It leads to the response to the external magnetic field positive both for three QDs. But with increasing interdot coupling strength, the three QDs are tending to generate a coherence bonding state and the spin of the adjacent QDs are beginning to form an antiparallel arrangement, as sketched in the lower part of figure 1(a). When the interdot coupling strength $t \geq 0.16\mathrm{meV}$, the QD-2 exhibits the negative value to the external magnetic field (figure 2(a)). Here, the reappearing Kondo effect accompanied with three-peak structure of the spectral function arises from the TQD system. The negative local susceptibility of QD-2 becomes an indirect proof of the coherence bonding state ($|\uparrow,\downarrow,\uparrow \rangle$ or $|\downarrow,\uparrow,\downarrow\rangle$) of TQD system, with the emerging of the reappearing Kondo effect. It is needed to pay more attention that only the result of susceptibility of the TQD system can be measured experimentally, but the local susceptibility of each QD can not acquired directly by experiment observation. Here, we show the variation of the local susceptibility on each QD only for the analysis of such Kondo physics.

We also analyze the variation of the temperature dependent susceptibility with different interdot coupling strengths $t$ (figure 2(b)). Several features in these curves are noteworthy. We focus first in high temperature case, the susceptibility of the TQD syetem is fitted well by a Curie-Weiss law $\chi = C/(T + \theta)$, where $C$ is a Curie constant, $T$ is the temperature and $\theta$ is a constant with a value in the thermal energy range ($0 < \theta < 300K$) \cite{1993cambrige}. We find the slopes of $1/\chi - K_{B}T$ curves at various interdot coupling strengths are almost the same value for high temperature. For weak interdot coupling strength, the susceptibility of the TQD system always shows the $T^{-1}$ temperature dependence. It is because that at weak coupling (especially if the temperature is much larger than the coupling energy), the response to the external magnetic field is positive both for the three QDs. More importantly, we focus on the low temperatures behavior of the susceptibility. The slopes of $1/\chi$ deviates from straight line at low temperature for strong interdot coupling strength $t$, where the local susceptibility $\chi$ of QD-2 changes progressively from positive value into the negative one (not shown here), under which, the TQD system gets into the reappearing Kondo regime. It provides another framework to study the reappearance of Kondo effect of the TQD system.

The susceptibility of the TQD system for very low temperature ($T \sim 0$) is not presented in our work. It is because that the HEOM method only studies the case of finite temperature but cannot deal with zero temperature case at present. The difficulty lies in the computational cost, which increases dramatically as the system temperature decreases. For a lower temperature, a higher truncation level is necessary to ensure numerical convergence, leading to a rapid growth of the required computational resources. It is, however, possible to design more efficient reservoir memory decomposition schemes to dramatically reduce the computational resources requirements\cite{2008jcp184112,2012prl266403,2015njp033009}.

The behavior of the reappearing Kondo effect at different temperatures is studied in figure 3(a), where we plot the results of the spectral functions $A(\omega)$ of QD-1(3) with different temperatures $K_{B}T$. To probe distinctly the reappearing Kondo effect of the TQD system, we adopt a large dot-lead coupling $\Delta = 0.3\mathrm{meV}$ and a strong interdot coupling $t = 0.25\mathrm{meV}$, the other parameters are the same as that in figure 1(b). The TQD system depicts a different varying behavior of Kondo effect from the single QD problem. Firstly, all the three splitting Kondo resonance peaks on QD-1(3) are robust at low temperatures. The reappearing Kondo effect also enhances with the low temperatures, leading to the increasing height of Kondo peaks for QD-1(3) with the decreasing temperature. For example, at the temperature $K_{B}T = 0.04\mathrm{meV}$, the reappearing Kondo peak ($\omega= 0$) for the TQD almost persevere in $0.225\mathrm{meV}^{-1}$, while it reaches to $0.245\mathrm{meV}^{-1}$ at a low temperature $K_{B}T = 0.03\mathrm{meV}$. It deserves special attention that the three peaks transfer to a broad packet with the increasing of the temperature ($K_{B}T= 0.15\mathrm{meV}$). Finally, all the Kondo peaks disappear at the temperature $T>T_{KT}$, here $T_{KT}$ is the Kondo temperature of the TQD system. We predict that the Kondo temperature $T_{KT}$ of TQD system is higher than single QD system, due to the fact that it needs more high temperature to quench the three splitting Kondo peaks of the TQD system's coherence bonding state.

\begin{figure}
\begin{minipage}[b]{0.5\textwidth}
\includegraphics[width=1\textwidth]{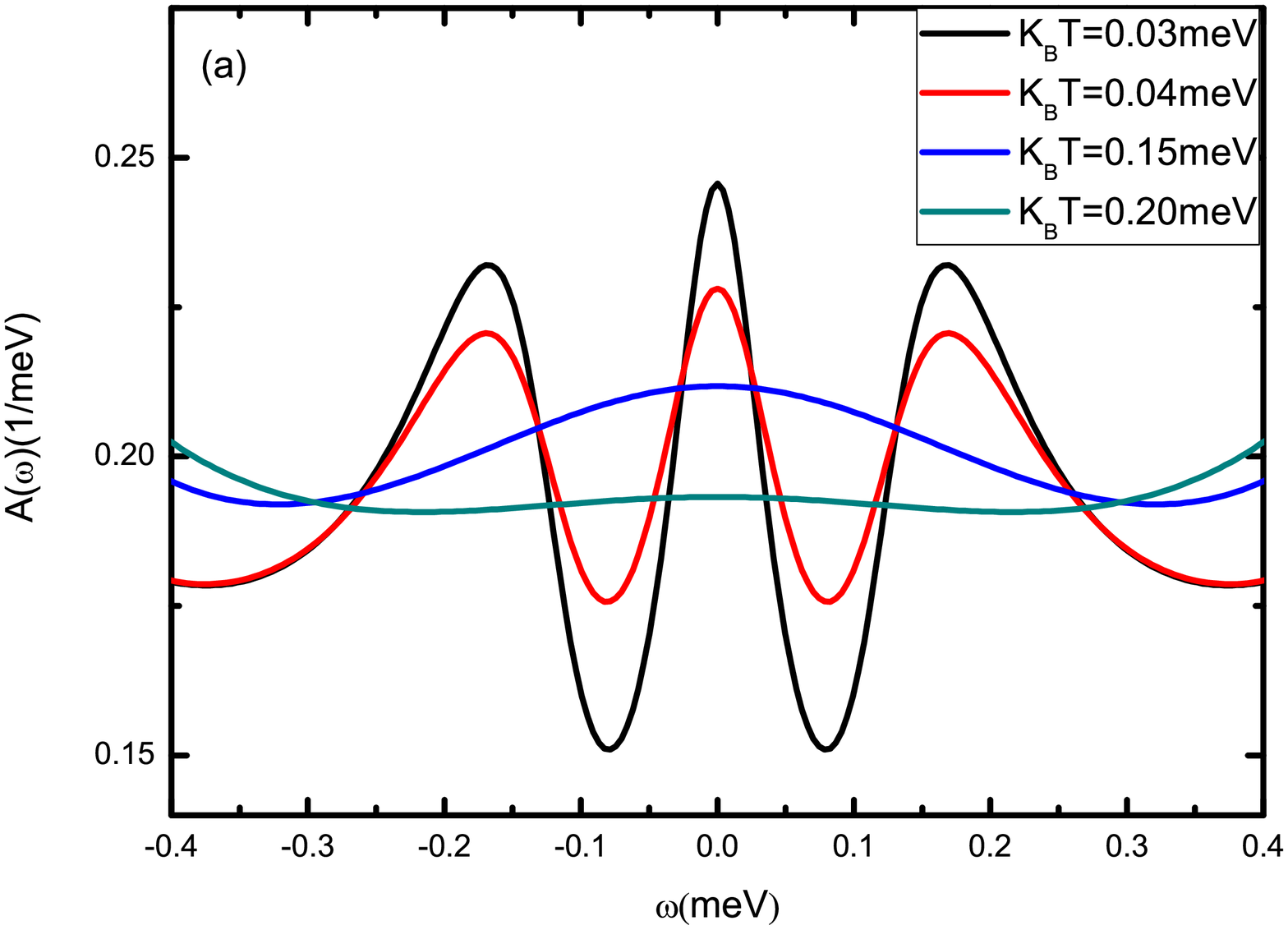}
\label{fig3(a)}
\end{minipage}%

\begin{minipage}[b]{0.5\textwidth}
\includegraphics[width=1\textwidth]{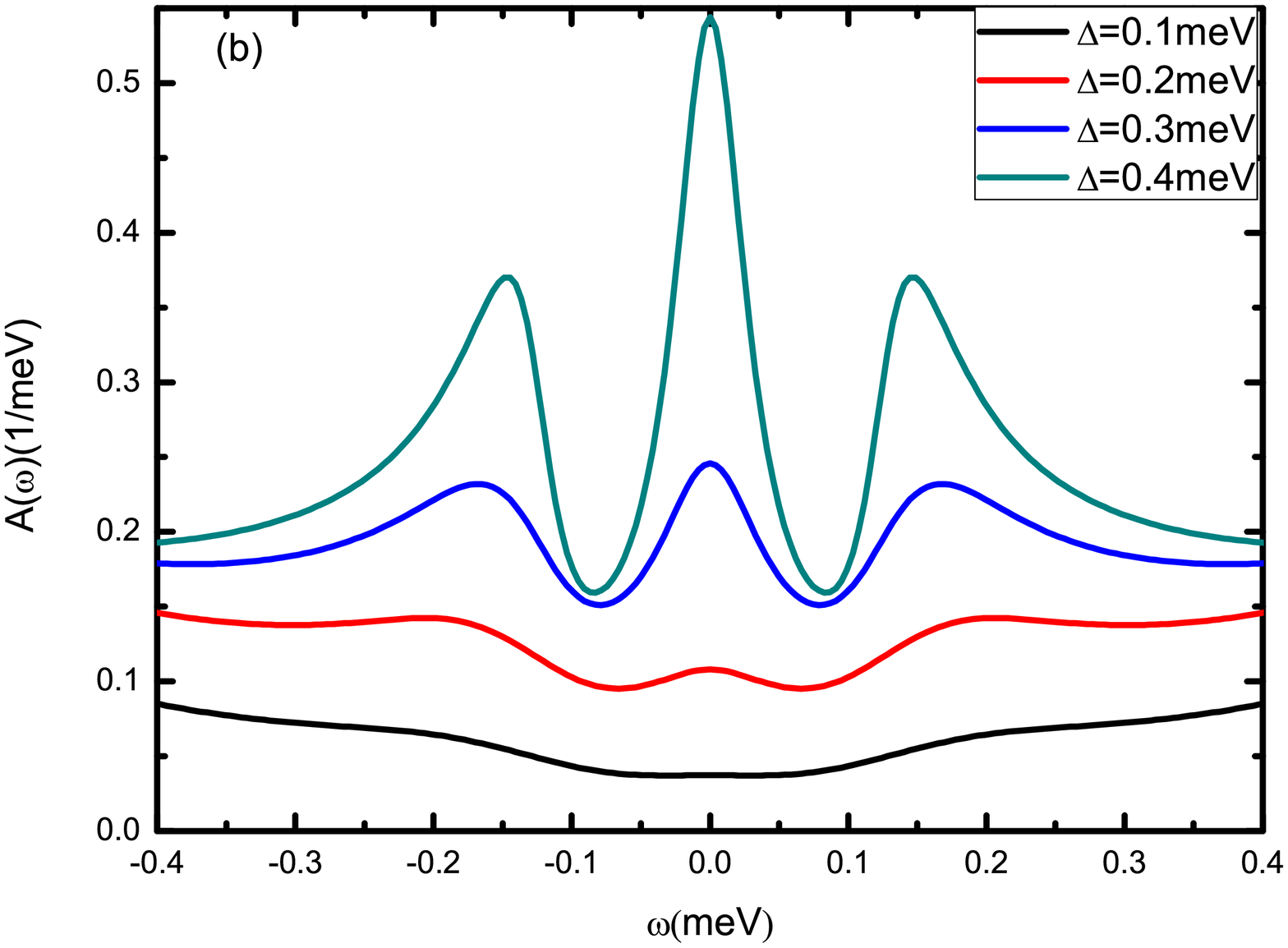}

\label{fig3(b)}
\end{minipage}
\caption{(Color online) (a) The spectral function $A(\omega)$ of the TQD system with different temperatures $K_{B}T$ for QD-1(3).The parameters adopted are $\epsilon_{i}(i=1,2,3) = -1.0\mathrm{meV}$ and $U_{i}(i=1,2,3) = 2.0\mathrm{meV}$, $W = 2.0\mathrm{meV}$, $t = 0.25\mathrm{meV}$, $\Delta = 0.3\mathrm{meV}$. (b) The spectral function $A(\omega)$ of the TQD system with different quantum dot-lead coupling strengths $\Delta$  for QD-1(3). The parameters adopted are $\epsilon_{i} (i=1,2,3)= -1.0\mathrm{meV}$ and $U_{i}(i=1,2,3) = 2.0\mathrm{meV}$, $W = 2.0\mathrm{meV}$, $t = 0.25\mathrm{meV}$, $K_{B}T = 0.03\mathrm{meV}$.}
\label{fig3}
\end{figure}

We then examine the dot-lead coupling strength $\Delta$ dependent reappearing Kondo effect. Figure 3(b) shows the spectral functions $A(\omega)$ of the QD-1(3) for interdot coupling strength $t = 0.25\mathrm{meV}$ case. We find an essential picture that the height of the reappearing Kondo peak rises rapidly with the increasing dot-lead coupling strength $\Delta$. We can attribute this transformation to the aforementioned increase of the effective Kondo temperature as a function of the dot-lead coupling. The mechanism can be understood via the Kondo physics of the single QD system. According to the analytical expression for Kondo temperature $T_{K}=\sqrt{\frac{U\tilde{\Delta}}{2}}e^{-\pi U/8\tilde{\Delta}+\pi\tilde{\Delta}/2U}$ \cite{1993cambrige} ($\tilde{\Delta}=2\Delta$ as two leads in our system), the Kondo temperature $T_{K}$ increases with the dot-lead coupling strength $\Delta$ with augmenting the height of the Kondo peak. Since the temperature of TQD system is fixed ($K_{B}T = 0.03\mathrm{meV}$), lead to the Kondo effect enhancing with the increase of $\Delta$, accompanied with the rising height of Kondo peak. For example, the height of the reappearing Kondo peak increases from $0.11\mathrm{meV}^{-1}$ at $\Delta = 0.2\mathrm{meV}$ to $0.54\mathrm{meV}^{-1}$ at $\Delta = 0.4\mathrm{meV}$ (see Figure 3(b)). Importantly, for weak interdot coupling strength $t$, the strong dot-lead coupling strength $\Delta$ only heighten the single Kondo peak, but can not develop the three-peak structure of the spectral function as shown in figure 3(b). Because the case that the reappearing Kondo effect origins from the coherence bonding state in the TQD system, which forms at a stronger interdot coupling strength $t$.

\begin{figure}
\includegraphics[width=0.95\columnwidth]{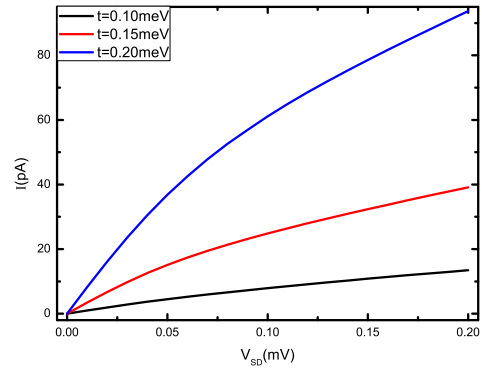}
\caption{(Color online) $I-V$ curves of the TQD system at various interdot coupling strengths $t$. The parameters adopted are $\epsilon_{i}(i=1,2,3) = -1.0\mathrm{meV}$ and $U_{i}(i=1,2,3) = 2.0\mathrm{meV}$, $W = 2.0\mathrm{meV}$, $\Delta = 0.2\mathrm{meV}$, $K_{B}T = 0.03\mathrm{meV}$.}
\label{fig4}
\end{figure}

Finally, we study the Kondo-correlated transport properties through the serially coupled TQD system. In this paper, the response current can be extracted from first-tier ($n = 1$) auxiliary density operators in HEOM space \cite{2012prl266403,2015njp033009}. The current through the TQD device for the appropriate values of voltages and interdot couplings, with the same parameters as figure 1(b) are plotted in figure 4. When the bias voltage is applied to the leads, the current flowing through the TQD system engenders. After the current rapidly increases, it reaches tardily a steady-state value. We find that the current increases strongly with the interdot coupling strength $t$. For example, at the bias voltage $V_{SD}=0.10 \mathrm{mV}$, the current is $I = 8 \mathrm{pA}$ for interdot coupling strength $t=0.10 \mathrm{meV}$, while it reaches $I = 60 \mathrm{pA}$ for the interdot coupling strength $t=0.20 \mathrm{meV}$. The Kondo-correlated transport behavior can be explained according to the Kondo effects picture of the TQD. The striking enhancement of the current is ascribed to the summational results of the first appearance and reappearing Kondo resonances in the TQD device. As a matter of fact, the observation of such a variation will provide a remarkable phenomenon of quantum coherence transport between the Kondo many-body states.

\section{CONCLUSIONS}

In summary, we have investigated the Kondo effect and spectral properties of the serially coupled TQD system based on the HEOM method. To depict the picture of this model, we consider the effective coupling between this TQD and the rest of the system. A clear picture of the reappearance of Kondo effect induced by the interdot coupling strength $t$ is described. This reappearing Kondo effects will develop a prominent transport behavior of the TQD system.

For weak interdot couplings, the conduction electrons of the lead screen the adjacent QD lead to the Kondo effect analogous to a single QD system. With the increase of interdot coupling strength $t$, the TQD system asymptotically transforms from the Kondo singlet state of individual QD to the coherence bonding state generated among the three QDs. So a crossover translation from local Kondo screening of the constituents to the formation of a local moment phase leads to a reappearing Kondo effect accompanied with a three-peak structure of spectral function of QD-1(3) rising from this strong interdot coupling strength. The properties of the susceptibility according to the interdot coupling strength $t$ and temperature $T$ provide another framework to study the reappearing Kondo physics of the TQD system. The transformation of the local susceptibility on QD-2 also confirm that the reappearing Kondo effect originates from the coherence bonding state of TQD system.

This Kondo effect of the TQD system on various factors is studied in detail. The reappearing Kondo effect exhibits a different temperature dependent behaviour from single QD problem, and TQD system owns higher Kondo temperature $T_{KT}$ than single QD system. The stronger dot-lead coupling strength $\Delta$ can also increase the Kondo temperature of the TQD device, leading to the enhanced reappearing Kondo effect. As a summational results of the fist appearance and reappearing Kondo effects, the transport current through TQD increases monotonically with increasing $t$. The characteristic universal signatures of the TQD system in physical quantities may be observed in experiments and is a prerequisite for the understanding and design of more complex structures, such as Kondo lattices. We hope our work will inspire and encourage experimental investigations of Kondo physics in coupled three QDs and related systems.

\section{ACKNOWLEDGMENT}

This work was supported by the NSF of China (No.\,11374363) and the Research Funds of Renmin
University of China (Grant No. 11XNJ026).
%




\end{document}